\documentclass[twocolumn,showpacs,preprintnumbers,amsmath,amssymb]{revtex4}

\usepackage{graphicx}
\usepackage{dcolumn}
\usepackage{bm}

\begin{document}

\noindent

\preprint{}

\title{Quantization from Hamilton-Jacobi theory with a random constraint}

\author{Agung Budiyono}

\affiliation{Institute for the Physical and Chemical Research, RIKEN, 2-1 Hirosawa, Wako-shi, Saitama 351-0198, Japan}


\begin{abstract}  

We propose a method of quantization based on Hamilton-Jacobi theory in the presence of a random constraint due to the fluctuations of a set of hidden random variables. Given a Lagrangian, it reproduces the results of canonical quantization yet with a unique ordering of operators if the Lagrange multiplier that arises in the dynamical system with constraint can only take binary values $\pm\hbar/2$ with equal probability.

\end{abstract}

\maketitle

\section{Motivation}

In the previous work \cite{AgungSMQ1}, we have developed a new method of quantization of system of spin-less particles  based on a specific modification of the classical dynamics of ensemble of trajectories parameterized by an unbiased (hidden) random variable. The fluctuations of the random variable is then characterized by the Planck constant. Given a wide class of classical Hamiltonians, the quantization is done by assuming rules of replacement of (deterministic) c-number by (stochastically parameterized) c-number to be applied to the classical Hamilton-Jacobi and continuity equations generated by the classical Hamiltonian. Assuming a specific type of distribution of the random variable, the modified Hamilton-Jacobi and continuity equations are then written into the Schr\"odinger equation with a unique quantum Hamiltonian, thus is free from the operator ordering ambiguity of the canonical quantization \cite{Ali review}. Moreover, one can always identify an effective velocity field which turns out to be equal to the actual velocity field of the particles in pilot-wave theory \cite{pilot-wave}. 

In the present paper, we shall show that the rules of replacement heuristically postulated in Ref. \cite{AgungSMQ1} can be derived from the Hamilton-Jacobi theory with a random constraint. The fluctuations of the constraint is assumed due to the presence of some background fields whose detail interaction with the particles are not known. The existence of background fields is also assumed in the Nelson stochastic mechanics to explain the origin of quantum fluctuations as time symmetric Brownian motion \cite{Nelson stochastic mechanics}. We shall then show that the hidden random variable postulated in Ref. \cite{AgungSMQ1} can be identified as the Lagrange multiplier that arises in the corresponding dynamical system with random constraint. 
 
\section{Hamilton-Jacobi theory with random constraint\label{S2}}    

Let us first review the Hamilton-Jacobi theory of classical dynamics. For simplicity, below we shall discuss the case of system with finite degrees of freedom. Let us denote the corresponding Lagrangian of the system as $\underline{L}(q,\dot{q})$ where $q=\{q_i\}$, $i=1,2,\dots,$ runs for all degrees of freedom, is the configuration coordinate, $t$ is time and $\dot{q}\doteq dq/dt=\{\dot{q}_i\}$ is the velocity. The canonical conjugate momentum is defined as 
\begin{equation}
\underline{p}_i(q,\dot{q};t)\doteq\frac{\partial\underline{L}}{\partial\dot{q}_i}. 
\label{classical momentum}
\end{equation}
If the Lagrangian is not singular, $\mbox{det}(\partial^2\underline{L}/\partial\dot{q}_i\partial\dot{q}_j)\neq 0$, which is assumed to avoid unnecessary complication, then the above equation can be solved in term of $\dot{q}$ as
\begin{equation}
\dot{q}_i=\dot{q}_i(q,\underline{p};t).
\label{classical velocity in term of momentum}
\end{equation}
Now let us define action as $\underline{I}\doteq\int\underline{L}(q,\dot{q})dt$. The actual trajectory connecting two spacetime points is then obtained by extremizing $\underline{I}$ with respect to variations of $(q,\dot{q})$ for a pair of fixed ends $\delta\underline{I}=\delta\int_{q_1(t_1)}^{q_2(t_2)}\underline{L}(q,\dot{q})dt=0$. This is the Hamilton's principle of stationary action which leads directly to the Euler-Lagrange and Hamilton equations. 

Another way to solve the above variational problem is through the Hamilton-Jacobi method \cite{Rund book}. First we construct a one-parameter family of hypersurfaces $\underline{S}(q;t)=\underline{\tau}$, where $\underline{\tau}$ is some parameter labeling the hypersurfaces, so that any point in the configuration space belongs only to one of the surfaces. Moreover a trajectory crosses each hypersurface only once and is nowhere tangent to it. $\underline{\tau}$ is thus a function of time, $\underline{\tau}=\underline{\tau}(t)$, so that one has 
\begin{equation}
\underline{\Delta}\doteq \frac{d\underline{\tau}}{dt}=\frac{d\underline{S}}{dt}=\partial_t\underline{S}+\dot{q}\cdot\partial_q\underline{S}. 
\label{Hamilton's function time derivative}
\end{equation}
By construction we have $\underline{\Delta}\neq 0$. The Hamilton-Jacobi method to get the actual trajectory then proceeds as follows. Let us consider two hypersurfaces $\underline{S}=\underline{\tau}$ and $\underline{S}=\underline{\tau}+d\underline{\tau}$, separated by infinitesimal $d\underline{\tau}$. Then given any point on  the hypersurface $\underline{S}=\underline{\tau}$, the actual trajectory that passes through this point and reaches the surface of $\underline{S}=\underline{\tau}+d\underline{\tau}$ is the one with a velocity $\dot{q}$ that minimizes $d\underline{I}/d\underline{\tau}$ \cite{Rund book}. Hence, the actual velocity must solve the following necessary Hamilton-Jacobi condition: 
\begin{equation}
\frac{\partial}{\partial\dot{q}_i}\Big(\frac{d\underline{I}}{d\underline{\tau}}\Big)=0, 
\label{H-J method}
\end{equation}
with fixed $d\underline{\tau}$. Since $d\underline{I}/d\underline{\tau}=\underline{L}/\underline{\Delta}$ and $\underline{\Delta}\neq 0$, Eq. (\ref{H-J method}) then reduces into   
\begin{equation}
\frac{\partial\underline{L}}{\partial\dot{q}_i}=\frac{\underline{L}}{\underline{\Delta}}\partial_{q_i}\underline{S}, 
\label{Hamilton's principle Jacobi reformulation} 
\end{equation}
where we have used $\partial\underline{\Delta}/\partial\dot{q}_i=\partial_{q_i}\underline{S}$ from Eq. (\ref{Hamilton's function time derivative}). 

Next, assuming that $\underline{L}/\underline{\Delta}=\theta(\underline{\tau})$, where $\theta$ is a function only of $\underline{\tau}$, then one can reparametrize $\underline{S}$ so that $\underline{L}=\underline{\Delta}$ \cite{Rund book}. In this case, one gets 
\begin{equation}
\underline{p}_i=\frac{\partial\underline{L}}{\partial\dot{q}_i}=\partial_{q_i}\underline{S}. 
\label{H-J condition} 
\end{equation}
Inserting this into Eq. (\ref{classical velocity in term of momentum}) one thus has 
\begin{equation}
\dot{q}_i=\dot{q}_i(q,\partial_q\underline{S};t).   
\label{velocity vs Hamilton's function}
\end{equation} 
Finally, substituting Eq. (\ref{velocity vs Hamilton's function}) back into Eq. (\ref{Hamilton's function time derivative}), recalling $\underline{L}=\underline{\Delta}$, one obtains the Hamilton-Jacobi equation
\begin{equation}
\partial_t\underline{S}+\dot{q}(q,\partial_q\underline{S};t)\cdot\partial_q\underline{S}-\underline{L}(q,\dot{q}(q,\partial_q\underline{S};t))=0.
\label{H-J equation}
\end{equation}

One can further show that the Hamilton-Jacobi equation of (\ref{H-J equation}) is equivalent to the Euler-Lagrange and the Hamilton equation \cite{Rund book}. In contrast to the latter two equations, the Hamilton-Jacobi equation describes a congruence of trajectories whose velocity field is given by Eq. (\ref{velocity vs Hamilton's function}). A single trajectory is then obtained if one also fixes the initial configuration of the system. Another important feature of Hamilton-Jacobi theory is that it imposes a local condition to the actual trajectory, that of Eq. (\ref{H-J method}). By contrast, Hamilton's principle gives a global condition to the actual trajectory. 

Now let us assume that the system under consideration depends on a set of hidden random variables $\xi\doteq(\xi_1,\xi_2,\dots)$ whose dynamical origin is not known. This for example might be due to the presence of background fields whose detail interaction with the particles is not known resulting in a stochastic motion of the latter. Single event is thus inherently random. Hence, one can only make prediction concerning an ensemble of copies of the system. Let us denote the joint-probability density of the fluctuations of $q$ and $\xi$ as $\Omega(q,\xi;t)$. The marginal probability densities of the configuration of the system $q$ and $\xi$ are then given as
\begin{equation}
\rho(q;t)\doteq\int d\xi \Omega(q,\xi;t)\hspace{2mm}\&\hspace{2mm}P(\xi)\doteq\int dq\Omega(q,\xi;t). 
\end{equation}
Here we have assumed that the probability density of $\xi$ is independent of time. Let us further assume that there is a random velocity field $v(q,\xi;t)$, of the same dimension as the system, so that $\Omega(q,\xi;t)$ has to satisfy the following differential equation: 
\begin{equation}
G(q,\dot{q};\Omega,v)=\frac{d(\ln\Omega)}{dt}+\partial_q\cdot v=\frac{\partial_t\Omega}{\Omega}+\frac{\partial_q\Omega}{\Omega}\cdot\dot{q}+\partial_q\cdot v=0. 
\label{quantum constraint}
\end{equation}
The above constraint might be interpreted that the velocity divergence gives the only source of (local in configuration space) change of entropy. 

Next let us construct a new family of hypersurfaces $S(q,\xi;t)=\tau$ so that for a fixed value of $\xi$, any point in configuration space belongs to only one of the hypersurfaces. Yet, a single point can belong to more than one hypersurfaces with different values of $\xi$. Namely $S(q,\xi;t)$ is now fluctuating due to the fluctuations of $\xi$. Further, let us assume that for a fixed value of $\xi$, a trajectory satisfying Eq. (\ref{quantum constraint}) crosses a hypersurface only once and nowhere tangent to it. $\tau$ is thus a function of $t$ and $\xi$, $\tau=\tau(t,\xi)$. Now, let us again consider two hypersurfaces $S=\tau$ and $S=\tau+d\tau$ separated by infinitesimal $d\tau$. Let us assume that within this interval, $\xi$ is fixed. Then, as in the case with no constraint discussed before, let us postulate that  given any point on  the hypersurface $S=\tau$, the actual trajectory that passes through this point and reaches the hypersurface $S=\tau+d\tau$ is the one with a velocity $\dot{q}$ that minimizes $d\underline{I}/d\tau$ and satisfying the constraint of Eq. (\ref{quantum constraint}). We thus have to solve Eq. (\ref{H-J method}) with the constraint of Eq. (\ref{quantum constraint}). Using the Lagrange method, this problem implies the following necessary condition:
\begin{equation}
\frac{\partial}{\partial\dot{q}_i}\Big(\frac{d\underline{I}}{d\tau}\Big)+\lambda(\xi)\frac{\partial G}{\partial\dot{q}_i}=0,
\label{constrained H-J condition}
\end{equation}
where $\lambda=\lambda(\xi)$ is the Lagrange multiplier. 

Notice that since the constraint is fluctuating randomly due to the fluctuations of $\xi$, then the Lagrange multiplier also depends on the value of the hidden variables thus is inherently random. Let us denote the probability density of $\lambda$ as $P(\lambda)$. As is clear from Eq. (\ref{constrained H-J condition}), $\xi$ appears explicitly in the equation only through $\lambda(\xi)$. Accordingly, we shall regard $\lambda$ as the effective hidden random variable and use it in place of $\xi$. Hence, for example, we shall write $\Omega(q,\lambda;t)$ instead of $\Omega(q,\xi;t)$ and so on. Further, by construction $\xi$ thus $\lambda$ in general depends on the configuration space and time. Later we shall assume that the  derivatives of $\lambda$ with respect to space and time are negligible as compared to that of $S$. The value of $\lambda$ can be obtained by inserting Eq. (\ref{constrained H-J condition}) back into Eq. (\ref{quantum constraint}). $\lambda$ thus depends on $\Omega$ and $v$. Below we shall go the other way around. Namely, we shall assume $\lambda$ with a specific statistical properties and look for a class of $\Omega$ and $v$ that satisfy Eqs. (\ref{quantum constraint}) and (\ref{constrained H-J condition}). 

One can then proceed as before to arrive at the following pair of equations \cite{Rund-book2}: 
\begin{eqnarray}
\partial_{q_i}S=\frac{\partial L}{\partial\dot{q}_i},\hspace{2mm} \partial_tS+\frac{\partial L}{\partial \dot{q}}\cdot\dot{q}-L=0. 
\label{pre}
\end{eqnarray}
where $L=L(q,\dot{q},\lambda;\Omega,v)$ is an extended Lagrangian defined as 
\begin{eqnarray}
L(q,\dot{q},\lambda;\Omega,v)\doteq\underline{L}(q,\dot{q})-\lambda G(q,\dot{q};\Omega,v)\nonumber\\
=\underline{L}(q,\dot{q};t)-\lambda\Big(\frac{\partial_t\Omega}{\Omega}+\frac{\partial_q\Omega}{\Omega}\cdot\dot{q}+\partial_q\cdot v\Big).
\label{quantum Lagrangian}
\end{eqnarray}
Equation (\ref{pre}) has to be solved together with the constraint of Eq. (\ref{quantum constraint}). From the left equation of (\ref{pre}), using Eq. (\ref{quantum Lagrangian}), one gets
\begin{equation}
p_i(q,\dot{q},\lambda;t,\Omega)\doteq\frac{\partial L}{\partial\dot{q}_i}=\underline{p}_i(q,\dot{q};t)-\lambda\frac{\partial_{q_i}\Omega}{\Omega}=\partial_{q_i}S. 
\label{fundamental equation 1}
\end{equation}
Again, assuming that the unconstrained Lagrangian $\underline{L}$ is not singular, the above equation can be solved in term of $\dot q$ to give
\begin{equation}
\dot{q}_i=\dot{q}_i(q,\underline{p};t)=\dot{q}(q,\partial_qS+\lambda\frac{\partial_q\Omega}{\Omega};t). 
\label{sub-quantum velocity}
\end{equation}
This velocity field implies the following continuity equation conserving the probability:
\begin{equation}
\partial_t\Omega+\partial_q\cdot\Big(\dot{q}(q,\partial_qS+\lambda\frac{\partial_q\Omega}{\Omega};t)\Omega\Big)=0.
\label{FPE} 
\end{equation}
Further, inserting Eq. (\ref{sub-quantum velocity}) into the right equation of (\ref{pre}), one obtains the following (modified) Hamilton-Jacobi equation:
\begin{eqnarray}
\partial_tS+\dot{q}(q,\partial_qS+\lambda\frac{\partial_q\Omega}{\Omega};t)\cdot\partial_qS\hspace{20mm}\nonumber\\
-L(q,\dot{q}(q,\partial_qS+\lambda\frac{\partial_q\Omega}{\Omega};t),\lambda;\Omega,v)=0, 
\label{HJME} 
\end{eqnarray}
where we have used the left equation of (\ref{pre}). 

We have thus Eqs. (\ref{FPE}) and (\ref{HJME}) that have to be solved in term of $\Omega$ and $S$ subjected to the condition of Eq. (\ref{quantum constraint}). They are coupled to each other since $\dot{q}$ now depends on $S$ and $\Omega$. We shall show in the next section by taking a concrete example that Eqs. (\ref{FPE}) and (\ref{HJME}) can be combined together to give a single partial differential equation for $S$ and $\Omega$ \cite{HJ-C equations}. The combined equation has to be solved with the condition of Eq. (\ref{quantum constraint}). To do this, we thus have to express $v$ that appears in both equations as function of $S$ and $\Omega$. On the other hand, to be meaningful, the constraint has to be consistent with the dynamics. Namely, if initially $\Omega$ satisfies the constraint of Eq. (\ref{quantum constraint}) then it must be so for any time as $\Omega$ is evolved by the dynamics through Eq. (\ref{FPE}). This naturally implies that the random velocity field of the constraint $v$ has to be related to velocity field of the dynamics $\dot{q}$. 
 
Now let us assume that $v(q,\lambda;t)$ is given as  
\begin{equation}
v_i(q,\lambda;t)\doteq\frac{\dot{q}_i(q,\lambda;t)+\dot{q}_i(q,-\lambda;t)}{2}=v_i(q,-\lambda;t). 
\label{Schroedinger magic}
\end{equation}
$v(q,\lambda;t)$ is thus uniquely determined by the choice of unconstrained Lagrangian $\underline{L}$. Moreover, let us further assume a class of solutions satisfying the following symmetry relations: 
\begin{equation}
S(q,\lambda;t)=S(q,-\lambda;t)+S_0(\lambda)\hspace{2mm}\&\hspace{2mm}\Omega(q,\lambda;t)=\Omega(q,-\lambda;t), 
\label{quantum symmetry}
\end{equation}
where $S_0(\lambda)$ is independent of $q$ and $t$. The former can be done by choosing an appropriate parameterization of the hypersurfaces, namely $\tau(\lambda,t)=\tau(-\lambda,t)+S_0(\lambda)$. Moreover, the latter implies that $\lambda$ is an unbiased random variable 
\begin{equation}
P(\lambda)=\int dq\Omega(q,\lambda;t)=P(-\lambda). 
\label{unbiased random Lagrange multiplier}
\end{equation}

Let us show that the above choice of random velocity field $v$ generates a constraint that is consistent with the dynamics. To see this, first, notice that taking the case when $\lambda$ is positive in the constraint of Eq. (\ref{quantum constraint}) add to it the case when $\lambda$ is negative and divided by two, imposing Eqs. (\ref{Schroedinger magic}) and (\ref{quantum symmetry}), one gets
\begin{equation}
\partial_t\Omega+\partial_q\cdot(v\Omega)=0,
\label{constraint: conserving probability current}
\end{equation}
which is just a continuity equation. Hence, in this case, the constraint is just a probability conservation equation generated by random velocity field $v$. On the other hand, from Eq. (\ref{FPE}), taking the case when $\lambda$ is positive add to it the case when $\lambda$ is negative and divided by two one gets, by virtue of Eqs. (\ref{Schroedinger magic}) and (\ref{quantum symmetry}),
\begin{equation}
\partial_t\Omega+\partial_q\cdot(v\Omega)=0, 
\end{equation}
which is the same as Eq. (\ref{constraint: conserving probability current}). Hence the constraint is consistent with the dynamics, as expected.  

\section{``effective'' pilot-wave model}

For simplicity, let us apply the general formalism developed in the previous section to ensemble of system of single particle of mass $m$ subjected to external potentials. The unconstrained (classical) Lagrangian then takes the form
\begin{equation}
\underline{L}(q,\dot{q})=\frac{1}{2}m\dot{q}^2+A(q)\cdot\dot{q}-V(q), 
\label{classical Lagrangian particle in potential}
\end{equation}
so that one has 
\begin{equation}
\underline{p}_i=m\dot{q}_i+A_i. 
\label{classical momentum particle in potential}
\end{equation}
The classical Hamiltonian reads
\begin{equation}
\underline{H}=\frac{1}{2m}(\underline{p}-A)^2+V. 
\label{classical Hamiltonian particle in potential}
\end{equation}

Inserting Eq. (\ref{classical momentum particle in potential}) into Eq. (\ref{fundamental equation 1}) one gets 
\begin{equation}
\dot{q}_i=\frac{\partial_{q_i}S}{m}-\frac{A_i}{m}+\frac{\lambda}{m}\frac{\partial_{q_i}\Omega}{\Omega}. 
\label{sub-quantum velocity particle in potential}
\end{equation}
Hence, Eq. (\ref{FPE}) becomes
\begin{equation}
\partial_t\Omega+\frac{1}{m}\partial_q\cdot\big((\partial_qS-A)\Omega\big)+\frac{\lambda}{m}\partial_q^2\Omega=0. 
\label{FPE particle in potential}
\end{equation}
Next, substituting Eq. (\ref{sub-quantum velocity particle in potential}) into Eq. (\ref{Schroedinger magic}) and imposing the assumption of Eq. (\ref{quantum symmetry}), $v$ is given by
\begin{equation}
v_i(q,\lambda;t)=\frac{1}{m}(\partial_{q_i}S-A_i). 
\label{quantum velocity particle in potential}
\end{equation}
Inserting Eqs. (\ref{quantum Lagrangian}), (\ref{classical Lagrangian particle in potential}), (\ref{sub-quantum velocity particle in potential}) and (\ref{quantum velocity particle in potential}) into Eq. (\ref{HJME}) one obtains 
\begin{eqnarray}
\partial_tS+\frac{(\partial_qS-A)^2}{2m}+V-\frac{2\lambda^2}{m}\frac{\partial_q^2R}{R}\hspace{20mm}\nonumber\\
+\frac{\lambda}{\Omega}\Big(\partial_t\Omega+\frac{1}{m}\partial_q\cdot\big((\partial_qS-A)\Omega\big)+\frac{\lambda}{m}\partial_q^2\Omega\Big)=0,
\label{preHJME particle in potential}
\end{eqnarray}
where we have defined $R\doteq\sqrt{\Omega}$ and used the following identity:
\begin{equation}
\frac{1}{4}\frac{\partial_{q_i}\Omega\partial_{q_j}\Omega}{\Omega^2}=\frac{1}{2}\frac{\partial_{q_i}\partial_{q_j}\Omega}{\Omega}-\frac{\partial_{q_i}\partial_{q_j}R}{R}. 
\label{fluctuation decomposition}
\end{equation} 
Substituting Eq. (\ref{FPE particle in potential}), Eq. (\ref{preHJME particle in potential}) then reduces into 
\begin{equation}
\partial_tS+\frac{(\partial_qS-A)^2}{2m}+V-\frac{2\lambda^2}{m}\frac{\partial_q^2R}{R}=0. 
\label{HJME particle in potential}
\end{equation}
Moreover, in this case, inserting Eq. (\ref{quantum velocity particle in potential}) into Eq. (\ref{constraint: conserving probability current}), the constraint then reads
\begin{equation}
\partial_t\Omega+\frac{1}{m}\partial_q\cdot\Big((\partial_qS-A)\Omega\Big)=0.
\label{general continuity equation}
\end{equation} 

We have thus pair of coupled Eqs. (\ref{HJME particle in potential}) and (\ref{general continuity equation}) parameterized by the fluctuating Lagrange multiplier $\lambda(\xi)$. Now, since $\lambda$ is non-vanishing, one can define the following complex-valued function:
\begin{equation}
\Psi(q,\lambda;t)\doteq R\exp\Big(\frac{i}{2|\lambda|}S\Big). 
\label{general wave function}
\end{equation}
It differs from the Madelung transformation in that $S$ is divided by $2|\lambda|$ instead of $\hbar$ so that one has $\rho(q;t)=\int d\lambda\Omega=\int d\lambda|\Psi|^2$. The pair of Eqs. (\ref{HJME particle in potential}) and (\ref{general continuity equation}) can then be recast into the following modified Schr\"odinger equation parameterized by the fluctuating Lagrange multiplier: 
\begin{equation}
i2|\lambda|\partial_t\Psi=\frac{1}{2m}(-i2|\lambda|\partial_q-A)^2\Psi+V\Psi,
\label{modified Schroedinger equation particle in potential}
\end{equation}
where we have imposed the assumption that the space and time derivatives of $\lambda$ are negligible as compared to that of $S$. 

Let us proceed to assume that $\Omega$ is factorisable as 
\begin{equation}
\Omega(q,\lambda;t)=\rho(q,|\lambda|;t)P(\lambda),
\label{separatos} 
\end{equation}
with $\int dq\rho(q,|\lambda|;t)=1$ for arbitrary value of $\lambda$. This guarantees that $\Omega$ is correctly normalized $\int d\lambda dq\Omega=1$. Moreover, let us assume that the Lagrange multiplier can only take binary values $\lambda(\xi)=\pm\hbar/2$ with equal probability 
\begin{equation}
P(\lambda)=\frac{1}{2}\delta(\lambda-\hbar/2)+\frac{1}{2}\delta(\lambda+\hbar/2). 
\label{unbiased binary random Lagrange multiplier}
\end{equation}
In this case, Eq. (\ref{modified Schroedinger equation particle in potential}) becomes 
\begin{equation}
i\hbar\partial_t\Psi_Q=\frac{1}{2m}(-i\hbar\partial_q-A)^2\Psi_Q+V\Psi_Q,
\label{Schroedinger equation particle in potential}
\end{equation}
where the wave function $\Psi_Q(q;t)$ is given by 
\begin{equation}
\Psi_Q(q;t)\doteq\sqrt{\rho(q,\hbar/2;t)}\exp\Big(\frac{i}{\hbar}S(q,\pm\hbar/2;t)\Big). 
\label{Schroedinger wave function}
\end{equation}
Hence $\rho(q;t)=|\Psi_Q(q;t)|^2$ holds by construction, and the phase is given by $S_Q(q;t)\doteq S(q,\pm\hbar/2;t)$. 

The quantum mechanical Schr\"odinger equation is thus reproduced as a specific case of the present statistical model when the Lagrange multiplier $\lambda$ is an unbiased binary random variable which can only take values $\lambda(\xi)=\pm\hbar/2$. To this end, it is interesting to mention Ref. \cite{Gaveau analytic continuation} which showed that the master equation of a particle moving with a fixed velocity, imposed to a random complete reverse of direction following a Poisson distribution, can be written into Dirac equation (in the same way that the Schr\"odinger equation is connected to the dynamics of Brownian motion) through analytic continuation. Note also that while $\lambda$ is a binary random variable, it is a function of the true hidden random variables $\xi$ which in turn may take continuous values. For example one may have $\lambda=\sqrt{\xi_1^2+\xi_2^2+\xi_3^2}=\pm\hbar$ so that $\xi$ lies on the surface of a ball of radius $\hbar$. If we divide the surface of the ball into two, attribute each devision $\pm\hbar$, respectively, and assume that $\xi$ moves sufficiently chaotic, then one will have $\lambda=\pm\hbar$ with equal probability.    

Moreover, in this case, the velocity field $v$ that generates the constraint then becomes    
\begin{eqnarray}
v_i(q,\pm\hbar/2;t)=\frac{1}{m}(\partial_{q_i}S(q,\pm\hbar/2;t)-A_i)\nonumber\\
=\frac{1}{m}(\partial_{q_i}S_Q(q;t)-A_i)\doteq \tilde{v}_i(q;t). 
\label{effective velocity particle in potential}
\end{eqnarray}
Keeping this in mind, Eqs. (\ref{constraint: conserving probability current}) or (\ref{general continuity equation}) then reads  
\begin{equation}
\partial_t\rho+\frac{1}{m}\partial_q\cdot\Big((\partial_qS_Q-A)\rho\Big)=0, 
\label{quantum continuity equation}
\end{equation}
where we have used Eq. (\ref{separatos}). It turns out that $\tilde{v}$ defined above is numerically equal to the actual velocity of the particle in pilot-wave theory \cite{pilot-wave}. This is also equal to the ``naively observable velocity field'' reported in Ref. \cite{Wiseman}, obtained using the notion of weak measurement \cite{AAV wm} within the standard interpretation of quantum mechanics.   

We have thus an ``effectively'' similar picture with pilot-wave theory in the sense that the particle always possesses definite position and momentum and further it moves ``as if'' it is guided by the wave function so that the ``effective velocity'' $\tilde{v}$ is given by Eq. (\ref{effective velocity particle in potential}). Hence, we can conclude that the statistical model developed in the present paper will reproduce the statistical wave-like interference pattern in slits experiment and tunneling over potential barrier \cite{Bohmian trajectory: simulation}. It is then tempting to further investigate in the future whether the model can lead to a description of quantum measurement which solves the infamous measurement problem. Note that to discuss the problem of measurement one needs to consider the time-irreversible process of registration which involves realistic description of apparatus and bath with large degrees of freedom \cite{ABN}.  

However, in contrast to pilot-wave theory, as is clearly shown in Eq. (\ref{effective velocity particle in potential}), $\tilde{v}$ is not the actual velocity of the particle, but the average of two actual velocities corresponding to Lagrange multiplier equal to $\hbar/2$ and $-\hbar/2$. Hence, while pilot-wave theory is strictly deterministic, the present model is inherently stochastic. Further, in this statistical model, rather than being postulated, the Schr\"odinger equation and a unique guidance relation emerge naturally by imposing a random constraint to the Hamilton-Jacobi condition. It is also evident that in contrast to pilot-wave theory, the wave function in the model is not physically real. It is just an artificial mathematical tool to describe the dynamics and statistics of the ensemble of trajectories, and  by construction the Born's statistics $\rho(q;t)=|\Psi_Q(q;t)|^2$ is valid for all time. The so-called quantum potential given by the last term of Eq. (\ref{HJME particle in potential}) with $\lambda=\pm\hbar/2$, which is argued by pilot-wave theory to be responsible for all peculiar quantum phenomena \cite{pilot-wave}, is generated by the local change in configuration space of the entropy part of the constraint. Note however that since the Schr\"odinger equation is time-reversal invariant then the total change of entropy should be vanishing. Finally while pilot-wave theory can deal with a single trajectory (since the wave function is assumed to be physically real satisfying the deterministic Schr\"odinger equation), the present model strictly concerns the dynamics of an ensemble of trajectories as in Nelson stochastic mechanics.  

\section{Quantization with unique ordering and direct physical interpretation}

We have shown in the previous section by taking an example of particle in external potentials how to develop from  a given classical Lagrangian a Schr\"odinger equation with unique quantum Hamiltonian. It can thus be regarded as to provide a method of quantization of classical system given its classical Hamiltonian. In contrast to the canonical quantization which is formal-mathematical, the method  presented in the previous section is based on Hamilton-Jacobi theory in the presence of random constraint so that the quantum-classical correspondence is physically kept transparent. 

To develop further formal comparison with canonical quantization, now let us extract some rules which can be applied directly given a classical Hamiltonian. First, let us define a scalar function $\underline{S}$ so that $d\underline{S}=\partial_t\underline{S}dt+\partial_q\underline{S}dq=\underline{L}dt$. Then, recalling that $dS=\partial_tS dt+\partial_qS dq=L dt$, from Eq. (\ref{quantum Lagrangian}), we have the following pair of relations:
\begin{eqnarray}
\partial_q\underline{S}=\partial_qS+\lambda\frac{\partial_q\Omega}{\Omega},\hspace{2mm}
\partial_t\underline{S}=\partial_tS+\lambda\frac{\partial_t\Omega}{\Omega}+\lambda\partial_q\cdot v. 
\label{fundamental equation}
\end{eqnarray}
On the other hand, the Hamilton-Jacobi equation of (\ref{HJME}) can be rewritten as 
\begin{equation}
\partial_tS+\lambda\frac{\partial_t\Omega}{\Omega}+\lambda\partial_q\cdot v+\dot{q}\cdot\Big(\partial_qS+\lambda\frac{\partial_q\Omega}{\Omega}\Big)-\underline{L}=0. 
\label{pre fundamental equation 1}
\end{equation}
Applying Eq. (\ref{fundamental equation}), the above equation thus becomes 
\begin{equation}
\partial_t\underline S+\underline{H}(q,\underline{p})|_{\underline{p}=\partial_q\underline{S}}=0,
\label{classical H-J equation}
\end{equation} 
where $\underline{H}(q,\underline{p})=\dot{q}\cdot\underline{p}-\underline{L}$ has the same form as classical Hamiltonian. Next, applying the left equation of (\ref{fundamental equation}), the velocity field of Eq. (\ref{sub-quantum velocity}) also becomes
\begin{equation}
\dot{q}=\dot{q}(q,\underline{p};t)|_{\underline{p}=\partial_q\underline{S}}=\frac{\partial\underline{H}}{\partial\underline{p}}\Big|_{\underline{p}=\partial_q\underline{S}},
\end{equation}
so that the continuity equation of Eq. (\ref{FPE}) now reads 
\begin{equation}
\partial_t\Omega+\partial_q\cdot\Big(\Omega\frac{\partial\underline{H}}{\partial\underline{p}}\Big|_{\underline{p}=\partial_q\underline{S}}\Big)=0.
\label{classical continuity equation} 
\end{equation}

Notice then that Eqs. (\ref{classical H-J equation}) and (\ref{classical continuity equation}) take the same form as the Hamilton-Jacobi equation and the continuity equation of classical mechanics given a classical Hamiltonian $\underline{H}(q,\underline{p})$. Hence, given the classical Hamiltonian, to get the Hamilton-Jacobi equation of (\ref{HJME}) and continuity equation of (\ref{FPE}) based on which we derive the Schr\"odinger equation, we can first develop the corresponding classical equations of (\ref{classical H-J equation}) and (\ref{classical continuity equation}), and apply the rules of Eq. (\ref{fundamental equation}). The remaining task is to express $v$ defined in Eq. (\ref{Schroedinger magic}) in term of the classical Hamiltonian $\underline{H}$. To do this, one can see that if the classical Hamiltonian is at-most-quadratic in  classical momentum, then $\dot{q}$ is a linear function of $\underline{p}=\partial_q\underline{S}=\partial_qS+\lambda(\partial_q\Omega/\Omega)$. In this case, evaluating Eq. (\ref{Schroedinger magic}) and taking into account Eq. (\ref{quantum symmetry}) one then obtains
\begin{eqnarray}
v=\frac{\dot{q}(q,\partial_qS+\lambda\frac{\partial_q\Omega}{\Omega};t)+\dot{q}(q,\partial_qS-\lambda\frac{\partial_q\Omega}{\Omega};t)}{2}\nonumber\\
=\dot{q}(q,\partial_qS;t)=\frac{\partial\underline{H}}{\partial\underline{p}}\Big|_{\underline{p}=\partial_qS}. 
\label{Schroedinger magic in classical Hamiltonian}
\end{eqnarray}
Inserting Eq. (\ref{Schroedinger magic in classical Hamiltonian}) into the constraint of Eq. (\ref{constraint: conserving probability current}), one thus has 
\begin{equation}
\partial_t\Omega+\partial_q\cdot\Big(\Omega\frac{\partial\underline{H}}{\partial\underline{p}}\Big|_{\underline{p}=\partial_qS}\Big)=0.
\label{quantum continuity equation 0} 
\end{equation}

It is then imperative to check whether the above equation is consistent with Eq. (\ref{classical continuity equation}). To see this, notice that $v$ in Eq. (\ref{Schroedinger magic in classical Hamiltonian}) is obtained by averaging $\dot{q}(\pm\lambda)$. Hence, Eq. (\ref{quantum continuity equation 0}) has to be obtained by averaging Eq. (\ref{classical continuity equation}) for the case of $\pm\lambda$ as well. One can evidently see by taking into account Eq. (\ref{quantum symmetry}) that for classical Hamiltonian at-most-quadratic in momentum, this is indeed the case. Namely, the constraint of Eq. (\ref{quantum continuity equation 0}) is indeed consistent with Eq. (\ref{classical continuity equation}). 

Given the classical Hamiltonian, we have thus two rules of Eqs. (\ref{fundamental equation}) and (\ref{Schroedinger magic in classical Hamiltonian}) to be applied to the classical mechanical equations of (\ref{classical H-J equation}) and (\ref{classical continuity equation}) and proceed in the way described in the previous section to arrive at the Schr\"odinger equation with a unique Hermitian quantum Hamiltonian. The above derived rules are just the rules of quantization proposed heuristically in Ref. \cite{AgungSMQ1}, that is Eq. (5) of Ref. \cite{AgungSMQ1}, where formal ``replacement'' there is re-interpreted in the present paper as physical ``substitution'' \cite{1/2}. Hence, we have given a justification of the rules postulated in Ref. \cite{AgungSMQ1} in term Hamilton-Jacobi condition with a random constraint. In other words, the hidden random variable $\lambda$ postulated in Ref. \cite{AgungSMQ1} is given physical interpretation as a random Lagrange multiplier that arises in Hamilton-Jacobi theory with a randomly fluctuating constraint.  
  
\section{Conclusion and discussion}

We have proposed a statistical model of quantization given a classical Lagrangian by assuming the existence of hidden random variables and accordingly imposing the Hamilton-Jacobi condition for the actual trajectory to a random constraint due to the fluctuations of hidden variables. Quantum fluctuations is shown to be emergent corresponding to a specific constraint depending uniquely on the choice of the classical Lagrangian and assuming that the Lagrange multiplier that arises in the dynamical system with constraint, which is fluctuating due to the fluctuations of the constraint, can only take binary values $\pm\hbar/2$ with equal probability. Given a classical system, the model leads to a unique quantum system with a straightforward physical interpretation.  

Let us mention several interesting problems that can be raised within the present statistical model. First, it is imperative to ask if the model can suggest new testable predictions beyond quantum mechanics. Such prediction, obtained by allowing fluctuations of $|\lambda|$ around $\hbar/2$ with very small yet finite width, is reported in Ref. \cite{AgungSMQ2}. Next, while we have offered an explanation on the physical origin of Planck constant in term of Lagrange multiplier, there is still a missing explanation on what determines its numerical value. Such a question of course beyond the standard quantum mechanics and is permissible only within a model in which quantum fluctuations is emergent \cite{Calogero conjecture}. Finally, notice that in the model, the random velocity constraint $v$ is uniquely determined by the choice of classical (unconstrained) Lagrangian. In this sense, the random constraint is already inherent in or self-generated by the system being constrained. Hence, to each quantum system there is {\it a hidden context} specific to the system.   

\begin{acknowledgments} 
The research is funded by the FPR program at RIKEN. 
\end{acknowledgments}


\begin{thebibliography}{10}
\bibitem{HJ-C equations} In general, the pair of Hamilton-Jacobi equation $\partial_tS+H(q,p)|_{\{p=\partial_qS\}}=0$ and the corresponding continuity equation $\partial_t\Omega+\partial_q\cdot(\Omega\frac{\partial H}{\partial p}|_{\{p=\partial_qS\}})=0$ can be developed by extremizing $\mathcal{I}=\int dtdq\Omega(\partial_tS+H(q,p)|_{\{p=\partial_qS\}})$ with respect to the variation of $\Omega$ and $S$ of fixed ends. In this sense, they are not independent to each other.
\bibitem{1/2} In Ref. \cite{AgungSMQ1} we have used a scale so that the Lagrange multiplier $\lambda$ in the present paper is related to $\lambda$ in that paper as $\lambda\mapsto\lambda/2$.
\end{thebibliography}
\end{document}